\documentclass{article}
\usepackage{spconf,amsmath,graphicx}
\usepackage{amssymb}
\usepackage[table]{xcolor}
\usepackage{booktabs}
\usepackage{multirow}
\usepackage{color}
\usepackage{hyperref}
\usepackage{url}

\def\bfX{{\mathbf X}}
\def\bfW{{\mathbf W}}

\newcommand{\parheader}[1]{{\smallskip \noindent \bf{#1}.}}

\title{JavaScript Convolutional Neural Networks for Keyword Spotting\\ in the Browser:\ An Experimental Analysis}
\name{Jaejun Lee \qquad Raphael Tang \qquad Jimmy Lin}
\address{David R. Cheriton School of Computer Science\\
University of Waterloo\\
\texttt{\{j474lee,r33tang,jimmylin\}@uwaterloo.ca}}
\begin{document}
%
\maketitle
\begin{abstract}
Used for simple commands recognition on devices from smart routers to mobile phones,
keyword spotting systems are everywhere. Ubiquitous as well are web applications, which
have grown in popularity and complexity over the last decade with significant
improvements in usability under cross-platform conditions.
However, despite their obvious advantage in natural language interaction, voice-enabled
web applications are still far and few between. In this work, we attempt to
bridge this gap by bringing keyword spotting capabilities directly into the browser.
To our knowledge, we are the first to demonstrate a fully-functional implementation of convolutional neural networks in pure JavaScript that runs in any standards-compliant browser.
We also apply network slimming, a model compression technique, to explore the accuracy--efficiency tradeoffs,
reporting latency measurements on a range of devices and software.
Overall, our robust, cross-device implementation for keyword spotting realizes a new paradigm for serving neural network applications,
and one of our slim models reduces latency by 66\% with a minimal decrease in accuracy of 4\% from 94\% to 90\%.

\end{abstract}
\section{Introduction}

With the rapid proliferation of voice-enabled devices, such as the Amazon Echo and Apple iPhone, speech recognition systems are
becoming increasingly prevalent in our daily lives. Importantly, these systems improve safety and convenience in hands-free interactions, such as 
using Apple's Siri to dial contacts while driving. However, a prominent drawback is that most of these systems perform speech recognition in the cloud, where 
a remote server receives all audio to be transcribed, as recorded by the device. Clearly, the privacy and security implications are significant:
the server may be accessed by other people---authorized or not. Thus, it is important
to capture the relevant speech only and not all incoming audio, all the while providing a hands-free experience.

Enter keyword spotting systems. They solve the aforementioned issues by implementing an on-device mechanism to ``wake up'' the intelligent agent, e.g., ``Okay, Google''
for triggering the Android assistant. This then allows the device to record and transmit a limited segment of relevant speech only, obviating the need
to be always-listening. Specifically, the task of keyword spotting (KWS) is to detect the presence of pre-specified phrases in a stream of audio, often with the end goal of 
wake-word detection or simple command recognition on device. Currently, state of the art uses lightweight neural networks~\cite{sainath2015convolutional, arik2017convolutional, 
tang2018deep, fernandez2018binarycmd}, which can perform inference in real-time even on low-end devices~\cite{fernandez2018binarycmd, tang2018experimental}.

Despite the popularity of voice-enabled products, web applications have yet to make use of keyword spotting. This is surprising, since modern
web applications are supported on billions of devices ranging from desktops to smartphones. Also, an in-browser KWS system would be able to perform the aforementioned 
simple commands recognition and wake-word detection. Thus, we attempt to close the gap between KWS systems and web applications in both research literature and industrial 
applications, building and evaluating such an in-browser system. Unfortunately, the browser is a highly inefficient platform for deploying neural networks, mainly due to 
poorly optimized matrix multiply routines. Fortunately, in recent years, the art of compressing neural networks has made significant advances in both general~\cite{liu2017learning, han2015deep, li2016pruning} and keyword spotting literature~\cite{fernandez2018binarycmd, zhang2017hello}. On our task, we demonstrate that network slimming~\cite{liu2017learning} is a simple yet highly effective method to achieve low latency with minimal impact on accuracy.

Thus, our main contributions are as follows: first, we develop a novel web application with an in-browser KWS system based on previous 
state-of-the-art~\cite{tang2018deep} models. Second, we provide the first set of comprehensive experimental results for the latency of an in-browser KWS system on a broad range 
of devices. Finally, to the best of our knowledge, we are the first to apply network slimming to examine various accuracy--efficiency operating points of a state-of-the-art KWS 
model. On the Google Speech Commands dataset~\cite{warden2018speech}, our most accurate in-browser model achieves an accuracy of 94\% while performing inference in less than 10 
milliseconds. With network slimming, we further reduce latency by 66\% while increasing the error rate by only 4\%.

\section{Background and Related Work}

\parheader{Keyword spotting}
KWS is the task of detecting a spoken phrase in audio, applicable to simple command recognition~\cite{tang2018deep, warden2018speech} and
wake-word detection~\cite{arik2017convolutional, sainath2015convolutional}. A typical requirement is that such a KWS system must be
small-footprint at inference time, since the target platforms are mobile phones, Internet-of-things (IoT) devices, and other portable electronics. To achieve
this goal, resource-efficient architectures using convolutional neural networks (CNNs)~\cite{tang2018deep, sainath2015convolutional} and recurrent neural networks
(RNNs)~\cite{arik2017convolutional} have been proposed, while other works make use of low-bitwidth weights~\cite{fernandez2018binarycmd, zhang2017hello}.
However, despite the pervasiveness of modern web browsers in devices from smartphones to desktops, and in spite of the availability of JavaScript-based deep learning toolkits,
implementing on-device KWS systems in web applications has never been done before.

\begin{figure}[t]
\centering
\includegraphics[scale=0.2,trim={0mm 0mm 0mm 3mm},clip]{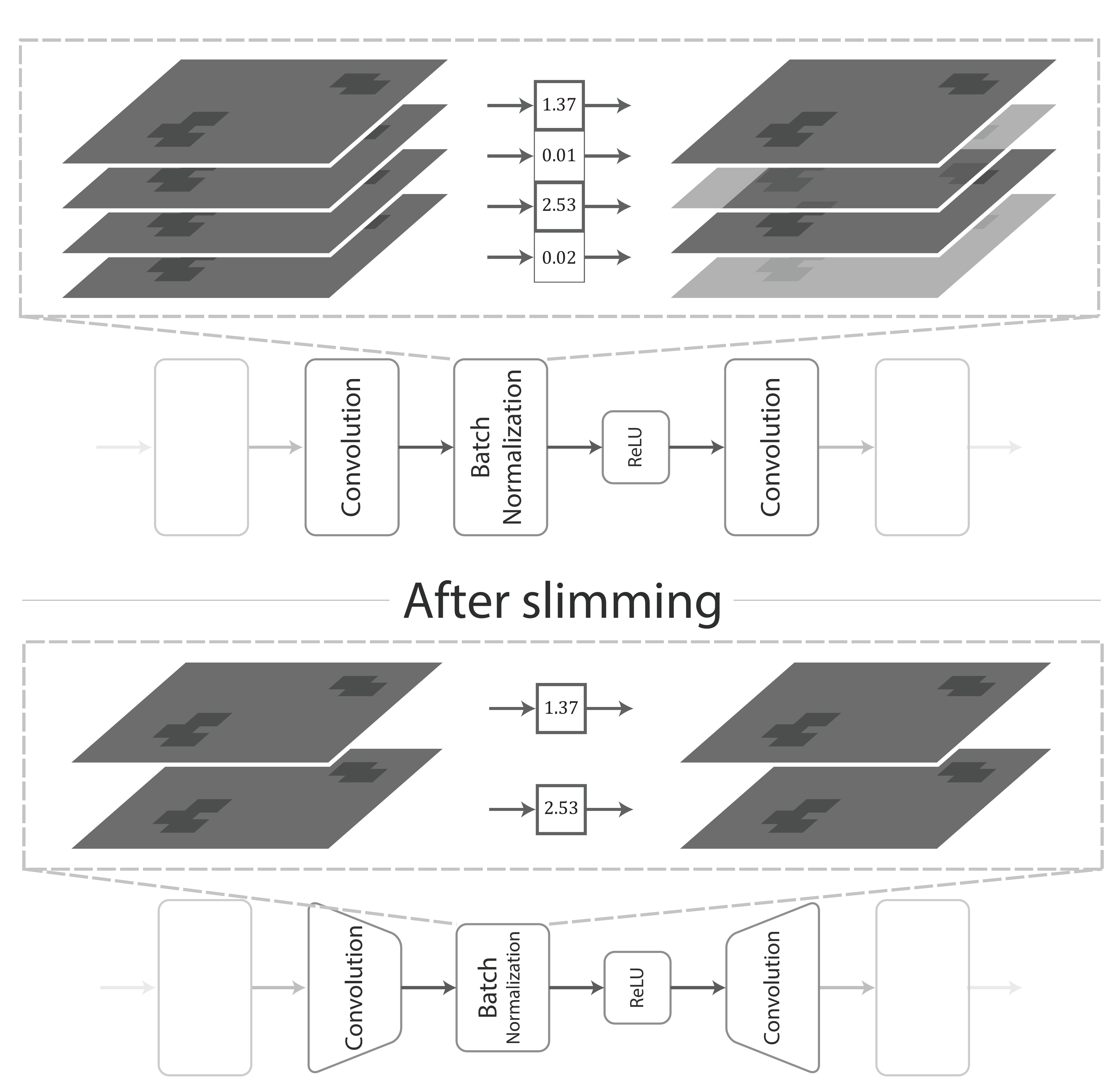}
\caption{An illustration of network slimming.}
\label{figure:slimming}
\end{figure}

\parheader{Compressing neural networks}
Sparse matrix storage leads to inefficient computation and storage in general-purpose hardware; thus, inducing structured sparsity in neural networks, e.g., on entire rows and
columns, has been the cornerstone of various compression techniques~\cite{liu2017learning, li2016pruning}. Network slimming~\cite{liu2017learning} is one such state-of-the-art approach
that have been applied successfully to CNNs: first, models are trained with
an $L_1$ penalty on the scale parameters in 2D batch normalization~\cite{ioffe2015batch} layers, which encourages entire channels to approach zero.
Then, a fixed percentage of smallest and hence unimportant scale parameters are removed, along with the correspondent preceding and succeeding filters in the convolution
layers (see Figure \ref{figure:slimming}). Finally, the entire network is fine-tuned on the training set---this entire process can optionally be repeated multiple times.

\section{Data and Implementation}
For consistency with past results~\cite{tang2018deep, tang2018experimental}, we train our models on the first version of the Google
Speech Commands dataset~\cite{warden2018speech}, which comprises a total of 65,000 spoken utterances for 30 short, one-second phrases.
To compare with past work~\cite{tang2018deep}, we pick the following twelve classes: ``yes,'' ``no,'' ``stop,'' ``go,'' ``left,''
``right,'' ``on,'' ``off,'' unknown, and silence.
It contains roughly 2,000 examples per class, including a few background noise samples of both man-made and artificial noise, e.g., washing dishes
and white noise. As is standard in speech processing literature, all audio is in 16-bit PCM, 16kHz mono-channel WAV format. We use the
standard 80\%, 10\%, and 10\% splits for the training, validation, and test sets, respectively~\cite{tang2018deep, warden2018speech}.

\subsection{Input preprocessing}
First, for dataset augmentation, the input is randomly mixed with additive noise from the background noise set~\cite{warden2018speech}---this helps
to decrease the generalization error~\cite{ko2015audio} and improve the robustness of the model under noisy conditions. Following the official TensorFlow implementation,
we also apply a random timeshift of Uniform$[-100, 100]$ milliseconds. Then, for the feature extraction step,
40-dimensional Mel-frequency cepstral coefficients (MFCCs) are computed, with a window size of 30 milliseconds and a frame shift of 10 milliseconds,
yielding a final preprocessed input size of $101 \times 40$ for each one-second audio sample.

\subsection{Model architecture}
\begin{figure}[t]
\centering
\includegraphics[scale=0.52]{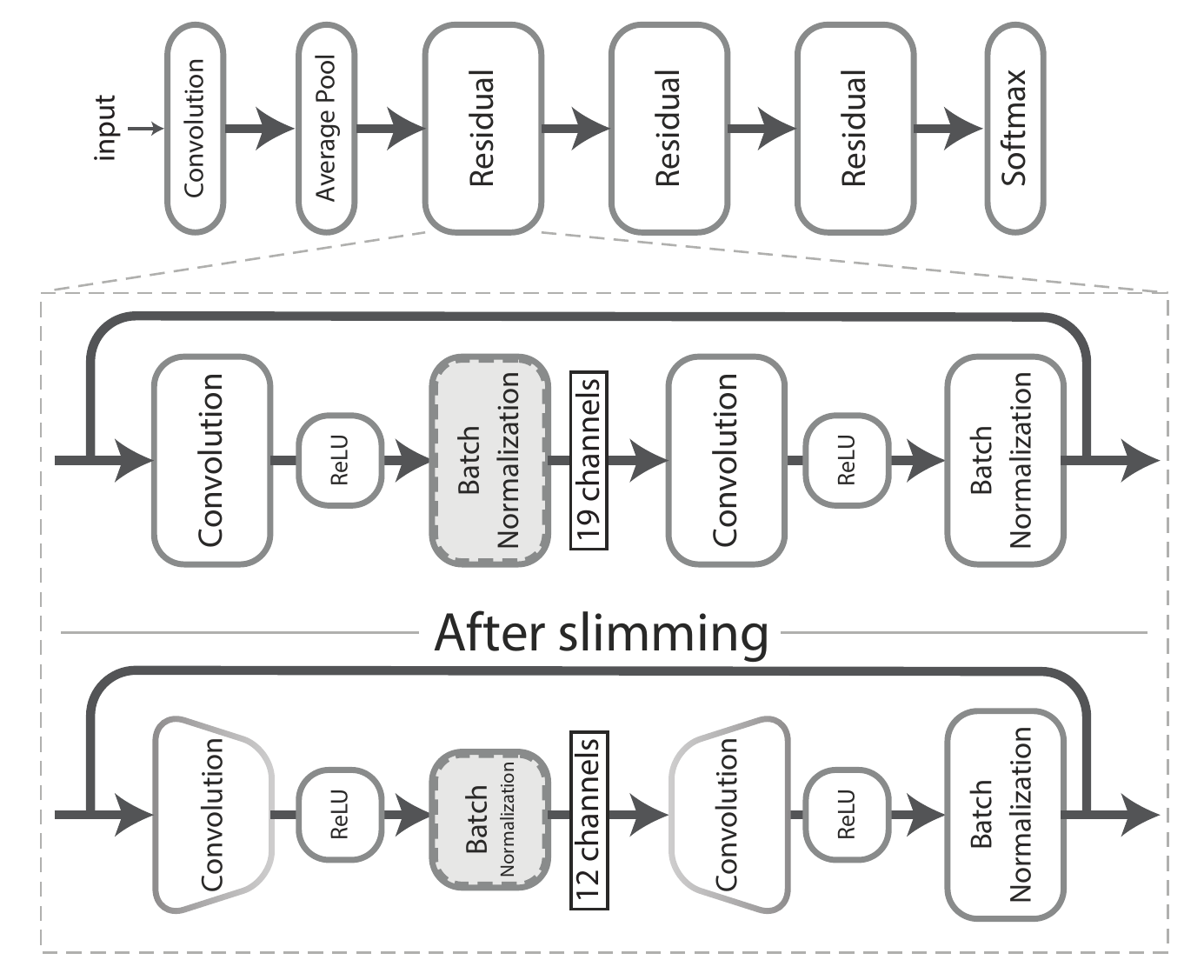}
\caption{Network slimming (40\%) of one residual block in \texttt{res8-narrow}, along with the full architecture.}
\label{figure:res_slimming}
\end{figure}

We use the \texttt{res8} and \texttt{res8-narrow} architectures from Tang and Lin~\cite{tang2018deep} as a starting point, which 
represent prior state of the art in residual CNNs~\cite{he2016deep} for KWS. In both models, given the input $\bfX \in \mathbb{R}^{101\times 40}$, we first
expand the input channel-wise by applying a 2D convolution layer with weights $\bfW \in \mathbb{R}^{C_{out} \times 1 \times (3 \times 3)}$ and padding of one on all sides.
This step results in an output of $\tilde{\bfX} \in \mathbb{R}^{C_{out} \times 101 \times 40}$, which we then downsample using an average pooling layer with a kernel
size of $(4, 3)$. Next, inspired by insights in image classification~\cite{he2016deep}, the output is passed through a series of three residual blocks
comprising convolution and batch normalization~\cite{ioffe2015batch} layers---Figure \ref{figure:res_slimming} illustrates one such block.
Finally, we average-pool across the channels and pass the features through a softmax layer across the twelve classes.

In the previous description, we are free to choose $C_{out}$ to dictate the expressiveness and computational footprint of the model. \texttt{res8} and \texttt{res8-narrow}
choose 45 and 19, respectively, for $C_{out}$. In total, \texttt{res8} contains 110K parameters and incurs 30 million multiplies per second of audio,
while \texttt{res8-narrow} uses 19.9K parameters and incurs 5.65 million multiplies per second.

\subsection{Implementation}
In-browser training of the model with JavaScript is not recommended due to poorly optimized computation routines such as matrix multiply and convolution.
Therefore, we use the official PyTorch model implementations\footnote{\url{https://github.com/castorini/honk}} at training time.
At inference time, weights are transferred from PyTorch to a web application implemented in TensorFlowJS.\footnote{\url{https://js.tensorflow.org/}}
Unlike Python, which is well-suited for developing audio processing applications~\cite{mcfee2015librosa}, in-browser JavaScript presents challenges in manipulating audio;
for example, many browsers restrict the sample rate of input audio to 44.1kHz only. To overcome these challenges, we use the Web Audio API for processing audio streams and
Meyda~\cite{rawlinson2015meyda} for computing MFCCs. The final values differ from MFCCs extracted by our LibROSA~\cite{mcfee2015librosa} Python back-end, even
with comprehensive patching; however, we found this to be a non-issue in evaluation.

Overall, we successfully enable KWS functionality in browser without any server-side inference. Since the audio data is quickly processed within the browser, it is much more efficient than transferring data over the network for inference. Furthermore, users are now freed from security and privacy implications, such as eavesdropping of network traffic and collection of personal speech data.

Since JavaScript does not guarantee same efficiency as Python with native C++ optimizations, we look for ways to further optimize in-browser inference. After exploring a number of options, we find that network slimming~\cite{liu2017learning} is a simple yet highly effective method to achieve this.

\parheader{Network slimming}
Since the compression technique~\cite{liu2017learning} relies on the presence of scale parameters in batch normalization layers, we cannot apply slimming as-is to the original
\texttt{res8-*}, which does not use affine transforms. For pruning, we must introduce a scale parameter $\gamma$ for each
batch normalization operation, corresponding to $\gamma \times \frac{\bfX - \mu}{\sigma}$ for input $\bfX$, mean $\mu$, and standard deviation $\sigma$.
Note that these new scale parameters are only introduced in the pruned architecture, because they are unnecessary for the full architecture. We create two configurations
of pruned models: one with 40\% of the parameters removed, and another with a more aggressive 80\% removed. We append \texttt{-40} and \texttt{-80}
to \texttt{res8} and \texttt{res8-narrow}, depending on the level of pruning.

\begin{table*}[t]
    \centering
    \small
    \begin{tabular}{rllccccc}
        \toprule[1pt]
        & \multirow{2}{*}{\raisebox{-3\heavyrulewidth}{\bf Device}} &
        \multirow{2}{*}{\raisebox{-3\heavyrulewidth}{\bf Processor}} &
        \multirow{2}{*}{\raisebox{-3\heavyrulewidth}{\bf Platform}} &
        \multicolumn{2}{c}{\bf \texttt{res8} } &
        \multicolumn{2}{c}{\bf \texttt{res8-narrow} } \\
        \cmidrule(lr){5-6}
        \cmidrule(lr){7-8}
        & & & & Latency (ms) & Accuracy (\%) & Latency (ms) & Accuracy (\%) \\
        \midrule
        \hspace{1mm}\multirow{5}{*}{GPU} & Desktop & GTX 1080 Ti & PyTorch  & 1 & 94.34 & 1 & 91.16 \\
        & Desktop & GTX 1080 Ti & Firefox & 8 & 94.06 & 7 & 90.91 \\
        & Macbook Pro (2017) & Intel Iris Plus 650 & Firefox  & 17 & 93.99 & 10 & 90.78 \\
        & Macbook Air (2013) & Intel HD 6000 & Firefox  & 34 & 93.99 & 19 & 90.78 \\
        & Galaxy S8 (2017) & Adreno 540 & Firefox & 60 & 94.06 & 43 & 88.96 \\
        \cline{1-8}\hspace{1mm}\multirow{6}{*}{CPU}
        & Desktop & i7-4790k (quad) & PyTorch  & 10 & 94.30 & 2 & 91.16 \\
        & Macbook Pro (2017) & i5-7287U (quad) & PyTorch  & 12 & 94.15 & 3 & 91.16 \\
        & Desktop & i7-4790k (quad) & Firefox  & 354 & 94.06 & 86 & 90.91 \\
        & Macbook Pro (2017) & i5-7287U (quad) & Firefox  & 338 & 93.99 & 94 & 90.78 \\
        & Macbook Air (2013) & i5-4260U (dual) & Firefox  & 485 & 93.99 & 115 & 90.78 \\
        & Galaxy S8 (2017) & Snapdragon 835 (octa) & Firefox & 1105 & 94.06 & 265 & 88.96 \\
        \bottomrule[1pt]
    \end{tabular}
    \caption{Latency and accuracy results on different platforms for the \texttt{res8} and \texttt{res8-narrow} models.}
    \label{table:lat_and_acc}
\end{table*}

\section{Evaluation}

Two main metrics for neural network application are accuracy and inference latency. To be consistent with the training process, the experiments use the same test set partitioned 
from the data. We conduct experiments and evaluate performance on desktop, laptop, and smartphone configurations to demonstrate the feasibility of our web application on a broad 
range of devices. First, we evaluate our application on a desktop with 16GB RAM, an i7-4790k CPU, and a GTX 1080 Ti. Then, we use the Macbook Pro (2017) and Macbook Air (2013) 
for our laptop configurations; the former has a quad-core i5-7287U CPU and an Intel Iris Plus 650 GPU, while the latter has a lighter dual-core i5-4260U CPU and an Intel HD 6000 
GPU. Finally, we choose the Galaxy S8 as our smartphone configuration. We select Firefox as the browser, and results are collected both with and without the 
existence of WebGL to evaluate the benefits of hardware acceleration.

\subsection{Results}

\parheader{In-browser KWS inference efficiency}
Measured with our university WiFi connection, the average latency to the Google server is about $\sim$25ms with standard deviation of 20ms. 
Network latency is much higher for transferring audio data. With a server written in Python, our evaluation presents an average latency of 481ms with standard deviation of 183ms 
for 16kHz mono-channel audio data. With in-browser inference, we achieve a serverless architecture which no longer suffers from variable network latency.

Table \ref{table:lat_and_acc} summarizes latency and accuracy results for both \texttt{res8} and \texttt{res8-} \texttt{narrow} on various devices. Note 
that results on our PyTorch implementation are included on laptop and desktop setups to compare to the standard baseline; the original implementation achieves an 
accuracy of 94.34\% for \texttt{res8} and 91.16\% for \texttt{res8-narrow} (see first few rows in the table). Slight differences are observed among platforms due to mismatch of MFCCs between LibROSA and Meyda. However, the accuracy for each model is consistent for every platform, confirming that our in-browser web application is indeed robust.

Even though latency is processor-dependent, the \texttt{res8-} \texttt{narrow} model performs inference in real-time on every platform, ranging from 7 to 43 milliseconds on GPU 
and 86 to 265 milliseconds on CPU configurations. Given that these delays are perceived by humans to be near-instantaneous~\cite{miller1968}, the latency we 
observe is sufficient for real-time interactive web applications, even considering the in-browser overhead. In fact, it is now 
feasible to deploy cross-platform neural network web applications even on mobile devices.

\begin{figure}[t]
\centering
\includegraphics[scale=0.58, trim={2cm 1cm 2cm 1.25cm}]{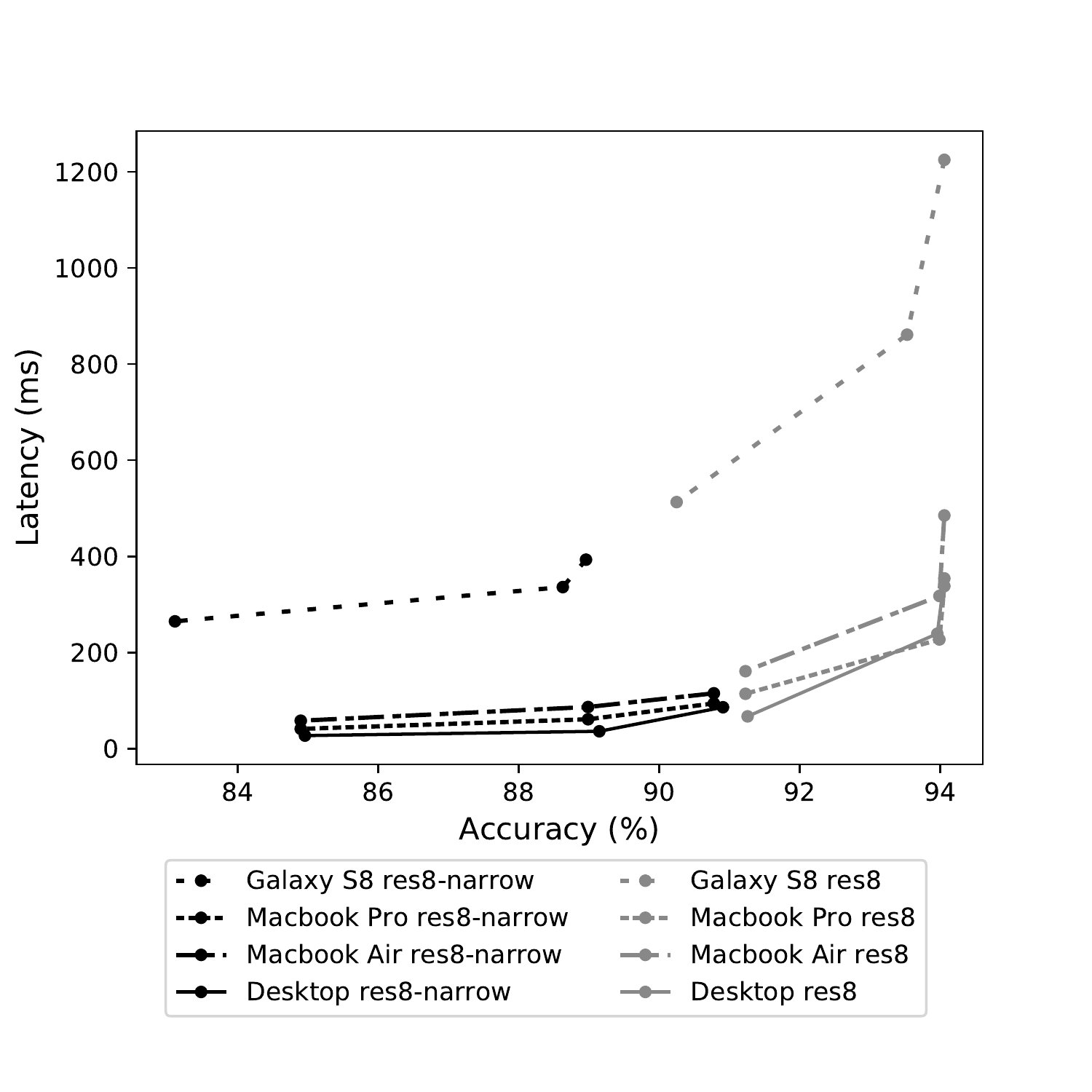}
\caption{Latency vs. accuracy curves.}
\label{figure:lat_vs_acc}
\end{figure}

\parheader{Latency--accuracy tradeoff}
Under the limited computational resources on mobile devices, network slimming can provide an option to tradeoff accuracy for inference latency.
To understand tradeoffs between latency and accuracy, we evaluate \texttt{res8} and \texttt{res8-narrow} models with 40\% and 80\% of its batch normalization layer pruned as well (see Figure \ref{figure:lat_vs_acc}); to illustrate the trend concisely, the figure includes results on CPU configurations only.

From \texttt{res8-narrow-80} to \texttt{res8-narrow-40}, accuracy increases by 4\% with minimal latency increase. However, starting from \texttt{res8-narrow-40}, the increase in 
latency is clear, indicating that obtaining higher accuracy comes at a cost. The slope of the curve increasingly steepens as accuracy increases, yielding 
tradeoff curves similar to those observed in other works~\cite{han2015deep, tang2018adaptive}.
Between \texttt{res8-40} and \texttt{res8}, change in accuracy is less than 1\% even though the most increase in latency is observed ranging from 111 ms to 363 ms. In other words, \texttt{res8-40} performs as well as \texttt{res8} while achieving lower latency.

Overall, we achieve a 50\% decrease in latency in \texttt{res8-} \texttt{narrow-80} and 66\% in \texttt{res8-80}, with only an absolute error rate increase of 4\%. \texttt{res8-narrow} on Macbook Pro requires 94 ms but drops down to 41 ms with \texttt{res8-narrow-80}. Similarly, latency on Galaxy S8 starts from 265 ms and decreases to 137 ms. Also, given that both accuracy and latency of \texttt{res8-narrow} are comparable to \texttt{res8-80}, network slimming provides option to replace one model to the other depending on target device.

\section{Conclusion}
In this paper, we realize a new paradigm for serving neural network applications by implementing KWS with in-browser inference. The serverless architecture allows our application to be efficient and cross-device compatible, with the additional benefit that users are freed from security and privacy implications, such as eavesdropping of network traffic and collection of personal speech data.

We implement a KWS web application that achieves an accuracy of 94\% while maintaining an inference latency of less than 10 ms on modern devices. With the goal of understanding 
accuracy and inference latency tradeoffs, we also analyze the impact of network slimming on existing \texttt{res8} and \texttt{res8-narrow} models. Our study shows that, with 
network slimming, our model yields a 66\% decrease in latency with a minimal increase in error rate of 4\%, along with accuracy--efficiency tradeoff curves like those 
observed in the past.


\end{document}